\documentclass[twocolumn,aps,prl,10pt,amsmath,amssymb,nofootinbib,showpacs,superscriptaddress,floatfix]{revtex4-1}

\DeclareFontFamily{U}{rcjhbltx}{}
\DeclareFontShape{U}{rcjhbltx}{m}{n}{<->rcjhbltx}{}
\DeclareSymbolFont{hebrewletters}{U}{rcjhbltx}{m}{n}

\usepackage{graphicx}
\usepackage{color}
\usepackage[usenames,dvipsnames]{xcolor}
\usepackage[colorlinks=true,linkcolor=Red,citecolor=Green,linktoc=page]{hyperref}
\usepackage{multirow}
\usepackage{float}
\usepackage{flushend}
\usepackage{balance}
\usepackage[varg]{txfonts}
\usepackage{ulem}
\usepackage{fancyhdr}

\DeclareMathSymbol{\lamed}{\mathord}{hebrewletters}{108}

\begin{document}
\title{Superinsulators, a toy realization of QCD in condensed matter}

\author{M.\,C.\,Diamantini}

\affiliation{NiPS Laboratory, INFN and Dipartimento di Fisica e Geologia, University of Perugia, via A. Pascoli, I-06100 Perugia, Italy}



\author{C.\,A.\,Trugenberger}

\affiliation{SwissScientific Technologies SA, rue du Rhone 59, CH-1204 Geneva, Switzerland}


\begin{abstract}
Superinsulators are dual superconductors, dissipationless magnetic monopole condensates with infinite resistance. The long-distance field theory of such states of matter is QED with dynamical matter coupled via a compact BF topological interaction. We will quantize the 2D model in the functional Schr\"odinger picture and show how strong entanglement of charges leads to a phase which is a single-color, asymptotically free version of QCD in which the infinite resistance is caused by the linear confinement of charges. This phase has been experimentally detected in TiN, NbTiN and InO  thin films, including signatures of asymptotically free behaviour and of the dual, electric Meissner effect. This makes superinsulators a ``toy realization" of QCD with Cooper pairs playing the role of quarks.

\end{abstract}
\maketitle


\section*{Introduction}
In (2+1) dimensions, gauge theories can be augmented by a topological term, the famed Chern-Simons (CS) term\,\cite{jackiw}. When this couples two different Abelian gauge fields, a vector and a pseudovector, one speaks of a mixed, or doubled CS term. In this paper we will consider an Abelian gauge model involving such a mixed CS term and the two corresponding Maxwell actions. This model is the (2+1)-dimensional version of topological BF models\,\cite{blau} in any number of dimensions. 

Such mixed CS and BF models were introduced as long-distance effective theories of condensed matter systems in\,\cite{dst}, where it was shown that they model the superconductor to insulator transition (SIT)\cite{sit} in Josephson junction arrays and thin superconducting films. Specifically, three phases were found when both gauge symmetries are compact, with gauge group U(1), and contain thus topological excitations describing vortices and point charges. First, the U(1) $\times $ U(1) phase with dilute topological excitations describes\,\cite{bm} the intermediate Bose metal phase\,\cite{das}, realizing a U(1) $\rtimes \ \mathbb{Z}_2^{\rm T}$ bosonic topological insulator\,\cite{lu}. When one of the two gauge symmetries is broken to $\mathbb{Z}$, instead, we have the superconductor and the superinsulator phases\,\cite{dst,si, vinokur, dtv1}, respectively. The latter is the subject of the present paper.

Superinsulators are a condensed matter realization of compact QED\,\cite{dtv1}, the simplest example of a strongly coupled gauge theory with a massive photon and linear confinement of charges\,\cite{polyakov}. While the pure gauge model, with only closed string excitations\,\cite{caselle} is non-renormalizable, it is known that coupling fermions, instead, does lead to a non-trivial fixed point of the renormalization group flow\,\cite{kleinert1}. The situation is, instead controversial in the compact Abelian Higgs model, one result pointing to the existence of a fixed point\,\cite{kleinert2}, another showing its absence\,\cite{lee}. Here we show that deep non-relativistic QED coupled to dynamical matter via a compact mixed Chern-Simons term has a Berezinskii-Kosterlitz-Thouless (BKT)\,\cite{bkt} fixed point separating an integer topological phase\,\cite{lu} from a confined phase. The integer topological phase\,\cite{lu} corresponds to a functional first Landau level and consists of an intertwined incompressible fluid of charges and vortices with a gap set by the CS mass. The confined phase is a highly entangled vortex condensate in which charges get bound to the ends of electric strings and the theory is asymptotically free, the BKT transition representing the strong-coupling infrared (IR) confinement phase. This is the superinsulation phase, realizing 't Hooft's old idea of quark confinement as dual superconductivity\,\cite{hooft}. Remarkably, this superinsulation phase has been experimentally observed in TiN, NbTiN and InO thin films\,\cite{vinokur}, including the explicit realization of confinement, asymptotic freedom and of the electric Meissner state\,\cite{electrostatics}. 

\section*{The model}
We consider a (2+1)-dimensional, non-relativistic model of dynamical matter coupled to electromagnetic gauge fields $A_{\mu}$ via a mixed CS term,
\begin{eqnarray}
S &&= \int dt\ d^2 x\ {-v\over 2e_0^2 } F_0 F^0+ {-1\over 2e_0^2 v} F_i F^i + {q \over 2\pi} A_{\mu} \epsilon^{\mu \alpha \nu} \partial_{\alpha} b_{\nu} 
\nonumber \\
&&+ {-v \over 2g_0^2} f_0 f^0 + {-1\over 2g_0^2 v} f_i f^i \ ,
\label{nonrelmodel}
\end{eqnarray}
where $e_0^2$ is the gauge coupling constant, with dimension [mass] in 2+1 dimensions (we use natural units $c=1$, $\hbar = 1$), and $v$ is the speed of light, a dimensionless number smaller than one in our units. Matter is formulated itself in terms of a fictitious pseudovector gauge field $b_{\mu}$ so that $j^{\mu} = (q/2\pi) f^{\mu}$, with $f^{\mu} = \epsilon^{\mu \alpha \nu} \partial_{\alpha} b_{\nu}$ the dual field strength, represents the conserved charge current. Correspondingly, $\phi^{\mu} =(1/2\pi) F^{\mu}$, $F^{\mu} = \epsilon^{\mu \alpha \nu} \partial_{\alpha} A_{\nu}$, is the vortex current. When the gauge symmetries are taken as compact, with radius $2\pi$ and  $2\pi/ q$ for $U(1)_b$ and $U(1)_A$, respectively, $q \in \mathbb{Z}$ plays the role of the charge quantum. The coupling $g_0^2$, also with dimension [mass], sets the energy scale of matter fluctutations. 

In applications to condensed matter systems, the relevant limit is the one in which the speed of light $v=1/\sqrt{\varepsilon \mu} <<1$ due to a very high dielectric permittivity $\varepsilon$ (while the magnetic susceptibility $\mu = O(1)$). We will thus consider the model (\ref{nonrelmodel}) in the limit $v \to 0$, in which only electric fields survive. This limit has been called the ``strong coupling limit" in\,\cite{grignani}; it is however, rather the deep non-relativistic limit (DNRL) and we will henceforth call it like that. The action is thus 
\begin{equation}
S = \int dtd^2 x\ {-1\over 2e^2 } F_i F^i + {q \over 2\pi} A_{\mu} \epsilon^{\mu \alpha \nu} \partial_{\alpha} b_{\nu} 
+ {-1\over 2g^2} f_i f^i \ ,
\label{strong}
\end{equation}
where we have reabsorbed the factor $v$ in a redefinition of the coupling constants, $e^2 = e_0^2 v$, $g^2 = g_0^2 v$. 

\section*{Functional Landau levels}

We shall quantize the model (\ref{strong}) in the functional Schr\"odinger picture. 
As usual for a gauge theory, the gauge components $A^0$ and $b^0$ are not dynamical fields, since they never appear with time derivatives. They are Lagrange multipliers, whose associated Gauss law constraints implement gauge invariance. They can be set to zero, $A^0=0$ and  $b^0 = 0$, after imposing the corresponding Gauss law constraints. This is called the Weyl gauge. 

The two canonical momenta conjugate to the canonical variables $A^i$ and $b^i$ are:
\begin{eqnarray} 
{\cal P}^i_A &&=  {\delta {\cal L} \over  \delta  (\partial_0 A^i )} = {1\over e^2}F^{0i} +  {q \over 4\pi }  \epsilon^{ij} b^j\ , \nonumber \\
{\cal P}^i_b &&= {\delta {\cal L} \over  \delta  (\partial_0 b^i )} = {1\over g^2}f^{0i} +  {q \over 4\pi }  \epsilon^{ij} A^j \ . 
\label{mc}
\end{eqnarray}
They are realized as functional derivatives, 
\begin{equation}
{\cal P}^i_A = -i {\delta \over \delta A^i} \ , \ \qquad {\cal P}^i_b = -i {\delta \over \delta b^i} \ .
\label{funder}
\end{equation}
The Hamiltonian density, when written in canonical variables takes the form 
\begin{equation}
{\cal H} = {e^2 \over 2} \left(\Pi^i_A\right) ^2 +  {g^2 \over 2} \left( \Pi^i_b \right)^2 \ ,
\label{hamden}
\end{equation}
where 
\begin{eqnarray}
\Pi^i_A &&= {\cal P}^i_A -{q \over 4\pi} \epsilon^{ij}b^j \ ,
\nonumber \\
\Pi^i_b &&= {\cal P}^i_b -{q\over 4\pi} \epsilon^{ij}A^j \ ,
\label{kinetic}
\end{eqnarray}
are the kinetic momenta. Due to the Chern-Simons term, the kinetic momenta do not commute,
\begin{equation}
\left[ \Pi^i_a ({\bf x}), \Pi^j_b ({\bf y}) \right] = -i \ {q\over 2\pi}\  \epsilon^{ij} \ \delta^2 ({\bf x} - {\bf y} ) \ .
\label{nocumm}
\end{equation}
which is tantamount to the presence of a non-trivial functional gauge connection 
\begin{eqnarray}
{\cal A}_A^i \left( A, b \right) = {q\over 4\pi} \epsilon^{ij}b_j \ ,
\nonumber \\
{\cal A}_b^i \left( A, b \right) = {q\over 4\pi} \epsilon^{ij}A_j \ ,
\label{fgc}
\end{eqnarray}
in the theory. Canonical momenta are not functional gauge invariant quantities anymore, since they can be traded with the connection by a functional gauge transformation of the wave functionals, 
\begin{eqnarray}
\Psi \left[ A^i, b^i \right] &&\to {\rm e}^{i {q\over 4\pi} \Lambda \left( A, b\right) } \ \Psi \left[ A^i, b^i \right] \ ,
\nonumber \\
\epsilon^{ij} b^j  &&\to \epsilon^{ij} b^j  + {\delta \over \delta A^i} \Lambda  \left( A, b\right) \ , 
\nonumber \\
\epsilon^{ij} A^j  &&\to \epsilon^{ij} A^j  + {\delta \over \delta b^i} \Lambda  \left( A, b\right) \ , 
\nonumber \\
\partial_i {\delta \over \delta A^i} \Lambda  \left( A^i, b^i \right) &&= 0 \ , \quad \partial_i {\delta \over \delta b^i} \Lambda  \left( A^i, b^i \right) = 0 \ ,
\label{gauge}
\end{eqnarray}
where the last conditions are required to respect traditional gauge invariance, encoded in the Gauss law constraints, see below. 
Only the electric fields $\Pi^i_A $ and the charge currents $\Pi^i_b$ are well-defined gauge invariant quantities.  

The functional connection (\ref{fgc}) is not pure gauge. The quantity
\begin{eqnarray} 
{\cal B}^{ij}({\bf x}-{\bf y})  &&=  {\delta \over \delta A^i ({\bf x})} {\cal A}_b^j \left(A({\bf y}), b({\bf y}) \right) - {\delta \over \delta b^j ({\bf y})} {\cal A}_A^i \left(A({\bf x}), b({\bf x}) \right) 
\nonumber \\
&&=  -{q\over 2\pi} \epsilon^{ij} \delta^2( {\bf x}-{\bf y}) \ ,
\label{magfield}
\end{eqnarray}
plays the role of a functional uniform magnetic field and appears as the commutator of kinetic momenta, exactly as in the traditional Landau problem of electrons in an external magnetic field. This shows that the Chern-Simons term plays the role of a functional magnetic field, i.e. of a non-trivial curvature in configuration space. 

Exactly as in the standard problem of Landau levels we can define lowering and raising operators 
\begin{eqnarray}
{\cal A}^i &&= \sqrt{\pi \over  q eg} \left( e \Pi^i_A - i g \epsilon^{ij} \Pi^j_b \right) \ ,
\nonumber \\
{{\cal A}^i}^{\dagger} &&= \sqrt{\pi \over q eg} \left( e \Pi^i_A + i g \epsilon^{ij} \Pi^j_b \right) \ ,
\label{lorai}
\end{eqnarray}
with commutation relation
\begin{equation}
\left[ {\cal A}^i({\bf x}), {{\cal A}^j}^{\dagger}({\bf y}) \right] = \delta^{ij} \ \delta^2 ({\bf x}-{\bf y}) \ .
\label{commlorai}
\end{equation}
In terms of these, the Hamiltonian takes the familiar form
\begin{equation}
H = m \sum_i \int d^2 {\bf x} \left( {{\cal A}^i}^{\dagger} ({\bf x}) {\cal A}^i({\bf x}) + {1\over 2} \ \delta^{ii} \ \delta^2( {\bf 0})\right)\ ,
\label{ham}
\end{equation}
where the second term represents the infinite ground state energy that has to be subtracted and $m=egq/2\pi$ is the topological energy gap. 
Finally, the Gauss law operators, implementing standard gauge invariance, are the constraints associated with the Lagrange 
multipliers $a_0$ and $b_0$, 
\begin{eqnarray} 
G_A \equiv \partial_i{\cal P}^i_A &+&  {q \over 4\pi } \partial_i \epsilon^{ij} b^j \ ,
\nonumber \\
G_b \equiv \partial_i{\cal P}^i_b &+&  {q \over 4\pi } \partial_i \epsilon^{ij} A^j \ .
\label{gl}
\end{eqnarray}
 At the quantum level these constraints must be imposed as conditions on physical states: 
\begin{equation} G_A  \Psi [A^i,b^i]   = 0 \ , \qquad   G_b  \Psi [A^i,b^i]  = 0  \ .
\label{glc}
\end{equation}
The ground state wave functional $\Psi_0$ is thus given by the symmetric gauge functional first Landau level, defined by 
${{\cal A}^i}({\bf x}) \Psi_0 [A^i,b^i]  = 0$, subject to the gauge constraints (\ref{glc}).

Localized excited states of unit norm and energy $m$ are created by the operators
\begin{eqnarray}
{\cal A}^{\dagger}_f = &&\int d^2 {\bf x} \ f( {\bf x}-{\bf x}_0) \ \hat n^i  {{\cal A}^i}^{\dagger} \ ,
\nonumber \\
&&\int d^2{\bf x} \ f^2({\bf x}- {\bf x}_0) = 1 \ ,
\label{excited}
\end{eqnarray}
with commutation relations 
\begin{eqnarray}
&&\left[ {\cal A}_f , {\cal A}^{\dagger}_f \right] = 1\ ,
\nonumber \\
&&\left[ H, {\cal A}^{\dagger}_f \right] \ \ = m {\cal A}^{\dagger}_f \ .
\label{exc}
\end{eqnarray}
These represent extended superpositions of matter and gauge fields. Their energy does not depend on the form factor $f( {\bf x}-{\bf x}_0)$, as long as it satisfies the normalization condition (\ref{excited}). 

In the original variables of (\ref{nonrelmodel}), the topological energy gap is $m = q e_0 g_0 v/2\pi $. In applications to condensed matter physics the relevant length scales are given by $1/e_0^2 = d/\alpha$ and $1/g_0^2 = \alpha \lambda^2/\pi^2 d$, where $\alpha $ is the fine structure constant, $\lambda$ is the London penetration depth of the superconducting material and $d$ is the film thickness, of the order of the coherence length. Therefore, the topological energy gap reduces to $m=(1/k) (v/d)$, where $k$ is the Ginzburg-Landau parameter of the material and we have considered $q=2$ for Cooper pairs. We shall consider the 2D non-relativistic limit $d\to 0$, $v\to 0$ so that $v/d$ is the highest frequency in the problem. Therefore, the higher Landau levels decouple and all relevant physics takes place in the lowest Landau level. 

Following \cite{jackiwbook, djt} we write the ground state functional as the product of a phase and a contribution that depends only on the transverse components of the two dynamical variables, $A^i_T$ and $b^i_T$: 
\begin{eqnarray}&\Psi_0 [ A^i,b^i ] =   {\rm e}^{  i \chi  \left( A^i,b^i \right)}  \ \Phi (A^i_T,b^i_T) \ , \nonumber \\
& \chi [A^i,b^i]  = {q \over 4 \pi} \int d^2{\bf x} \left( b {\partial_i \over \Delta  } A^i   + A {\partial_i \over \Delta  } b^i \right)  \ , \nonumber \\
& \Phi [A^i_T,b^i_T] = \exp   {-  q \over 4\pi} \int d^2{\bf x} \left( {g \over e} ( A^i_T )^2 + {e\over g} ( b^i_T  )^2  \right) \ , 
\label{gswf}
\end{eqnarray}
where  $A= \epsilon^{ij} \partial_i A^j$, $b = \epsilon^{ij} \partial_i b^j$, $\Delta = \partial_i \partial_i$ and 
$A^i_T  =  P^{ij}  A^j$, $b^i_T =  P^{ij}  b^j $, with the projector $ P^{ij} $ onto the transverse part of the gauge fields given by $ P^{ij} = \left (\delta^{ij} -  {\partial^i  \partial^j\over \Delta  } \right) $. 
Using the Hodge decomposition for the spatial components of the two gauge fields $A^i$ and $b^i$:
\begin{eqnarray}A^i &=& \partial_i \xi + \epsilon^{ij} \partial_j \phi \ , \nonumber \\
b^i &=& \partial_i \lambda + \epsilon^{ij} \partial_j \psi \ ,
\label{hod}
\end{eqnarray}
we can rewrite $\Psi_0 [A^i,b^i]$ as:
\begin{equation}
\Psi_0 [A^i,b^i]  = {\rm e}^ {{i q \over 4 \pi} \int d^2{\bf x} \ \left( \psi \Delta \xi + \phi \Delta \lambda \right)} 
{\rm e}^{{ - q \over 4 \pi} \int d^2{\bf x} \ \left(   \kappa (\partial_i \phi)^2 +  {1\over \kappa} (\partial_i \psi)^2\right)} \ ,
\label{ngs}
\end{equation}
where $\kappa = g/e$ represents the dimensionless coupling constant of the theory. 
As always in Chern-Simons gauge theories, gauge invariance is realized with a 1-cocycle\,\cite{jackiwbook}, which manifests itself in the phase in (\ref{gswf}) and (\ref{ngs}). This can be expressed also as
\begin{equation}
 {\rm e}^{i \chi  \left( A^i,b^i \right)}  = {\rm e}^{ {i\over 2}  \int d^2 {\bf x} \left( q\lambda \phi^0 + \xi j^0 \right)} \ .
 \label{phase}
 \end{equation}
 
Two possibilities have to be considered. In the simplest case both gauge symmetries are non-compact, with gauge group 
$\mathbb{R}$. In this case, neither charges nor vortices are quantized and ground state quantum correlation functions of their densities 
are given by
\begin{eqnarray}
\langle j^0({\bf x}) j^0({\bf y}) \rangle_c &&= {1\over 4\pi^2} {1\over Z_{\psi}} \int {\cal D}\psi \  \Delta \psi ({\bf x}) \Delta \psi ({\bf y}) \  e^{{-q  \over 2\pi \kappa} \int d^2 {\bf x} \  \left( \partial_i \psi \right)^2} \ ,
\nonumber \\
Z_{\phi} &&= \int {\cal D} \psi \ e^{{-q \over 2\pi \kappa } \int d^2 {\bf x} \ \left( \partial_i \psi \right)^2} \ ,
\label{correlation}
\end{eqnarray}
where the subscript ``$c$" denotes connected correlation functions and 
with an analogous expression for vortices in terms of the field $\phi$ and with $\kappa \to 1/\kappa$. 
This gives
\begin{equation}
\langle j^0({\bf x}) j^0({\bf y}) \rangle_c = {\pi \kappa\over q} \ \Delta \delta^2 ({\bf x} -{\bf y}) \ , 
\langle \phi^0({\bf x}) \phi^0({\bf y}) \rangle_c ={\pi \over q \kappa} \ \Delta \delta^2 ({\bf x} -{\bf y}) \ ,
\label{shortrange}
\end{equation}
which represent short-range, screened correlations. The vanishing screening length implied by these expressions is a consequence of the deep non-relativistic limit $v\to 0$ in (\ref{strong}). In the general case (\ref{nonrelmodel}) the real part of the ground state wave functional (\ref{gswf}) is modified to
\begin{equation} 
\Phi [A^i_T,b^i_T] = {\rm e}^{ -\int d^2{\bf x} \ {1\over 2e^2} A_{\rm T}^i \sqrt{ m^2 - v^2 \Delta }\  A_{\rm T}^i
+ {1\over 2g^2} b_{\rm T}^i \sqrt{ m^2 - v^2 \Delta } \ b_{\rm T}^i } \ ,
\label{modified}
\end{equation}
which shows that both charge- and vortex densities are correlated on a typical length $\lambda_{\rm corr} = v/m=\lambda$. 

In the Hamiltonian formalism, the operator that inserts external point charges $\pm Q$ at ${\bf x}_1$ and ${\bf x}_2$ is the exponential line integral\,\cite{sakita}, 
\begin{eqnarray}
&&\Psi_0^{\ \rho} [A^i,b^i] = {\rm e}^{ iQ \int_{{\bf x}_1}^{{\bf x}_2} A^i d{\bf x}^i } \Psi_0 [A^i,b^i]  \ ,
\nonumber \\
&&\rho = Q\  \delta^2 \left( {\bf x} -{\bf x}_1 \right) - Q\  \delta^2\left( {\bf x} -{\bf x}_2 \right) \ .
\label{ext}
\end{eqnarray}
Using the functional gauge invariance (\ref{gauge}) (without the last, zero divergence condition due to the presence of an external charge distribution), this can be reabsorbed by a simple shift 
\begin{equation}
\epsilon^{ij} b^j \to \epsilon^{ij}b^j + {4\pi Q\over q} {\partial_i \over \Delta} \rho \,
\label{shift}
\end{equation} 
which amounts to a corresponding shift
\begin{equation}
\Delta E_0 =  \int d^2 {\bf x} \ \rho {1\over \Delta} \rho \ ,
\label{log}
\end{equation}
in the ground state energy, showing that, in this phase, external probe charges interact logarithmically. Note that this is not in contradiction with short-range charge density correlation functions. It is only {\it external} probes that interact logarithmically, {\it dynamical} charges in the model are screened by the topological interactions. 

\section*{Confinement  and asymptotic freedom} 

Things change when the two gauge symmetries are compact U(1) instead of $\mathbb{R}$. In this case, the fields $\xi $ and $\lambda$ in (\ref{hod}) are angles and the identity $\epsilon^{ij}\partial_i \partial_j \theta = 2\pi \delta^2 ({\bf x})$, in polar coordinates $r, \theta$, implies the existence of quantized vortices and point charges.  As we now show, in this U(1) $\times$ U(1) case, the fields $\xi $ and $\lambda$ become new dynamical fields embedding the dynamics of these additional degrees of freedom.

We start by noting that, in case of compact gauge symmetries, the cocycle is changed to
\begin{equation}
 {\rm e}^{i \chi  \left( A^i,b^i \right)}  = {\rm e}^{ {i\over 2}  \int d^2 {\bf x} \left( q\lambda \phi^0 + \xi j^0 \right)} 
 {\rm e}^{i {q\over 2} \sum_i \int d^2{\bf x} \xi( {\bf x}) N( {\bf x}, {\bf x}_i N_i) +  {i\over 2} \sum_i \int d^2{\bf x} \lambda( {\bf x}) \Phi ( {\bf x}, {\bf x}_i \Phi_i)} \ ,
 \label{newco}
 \end{equation}
 where
 \begin{eqnarray}
 N \left( {\bf x}, {\bf x}_i ,N_i \right) = N_i \ \delta^2 \left( {\bf x}- {\bf x}_i \right) \ ,
 \nonumber \\
 \Phi \left( {\bf x}, {\bf x}_i ,\Phi_i \right) = \Phi_i \ \delta^2 \left( {\bf x}- {\bf x}_i \right) \ ,
 \label{point}
 \end{eqnarray}
represent the additional point particle and vortex degrees of freedom. The integers $N_i$ and $\Phi_i$ encode the particle and vortex numbers, while ${\bf x}_i$ denotes their locations. We shall now consider gauge sector observables in entangled mixed states in which the charge degrees of freedom $N \left( {\bf x}, {\bf x}_i ,N_i \right) $ and $\psi ({\bf x})$ are considered as the non-observed environment over which we trace.  The expectation values of gauge sector operators $O(\phi, \xi)$ in this mixed state are given by  
\begin{eqnarray}
&&\langle O \rangle ={1\over Z} \int {\cal D}\phi {\cal D}\xi {\cal D}\psi \ \sum_N {z^N\over N!} \sum_{{\bf x}_1 \dots {\bf x}_N} \sum_{N_1 \dots N_N = \pm 1} O(\phi, \xi) 
\nonumber \\
&&{\rm e}^{{iq\over 2} \sum_i \int d^2 {\bf x} \ \xi ({\bf x}) N \left( {\bf x}, {\bf x}_i N_i \right) +i \int d^2{\bf x} \ {q\over 4\pi}\xi \Delta \psi} 
{\rm e}^{{ - q \over 2 \pi} \int d^2{\bf x} \left(  \kappa (\partial_i \phi)^2 +  {1\over 4\kappa} (\partial_i \psi)^2\right)} \ .
\label{exp}
\end{eqnarray}
where $Z$ is the normalization factor and $\psi$ denotes the difference $\psi = \psi_{\rm bra} -\psi_{\rm ket}$ between bra and ket states. We have also used the dilute charge approximation in which only interferences between point charge states differing by one unit are taken into account. The quantum fugacity parameter $z$ governs the entanglement. For small $z$ we have a highly entangled state of charge degrees of freedom, for $z\to \infty$ charges are liberated as independent degrees of freedom. 

At this stage both the integration over the transverse field $\psi$ and the summation over charge interference configurations can be done explicitly\,\cite{polyakov}, with the result
\begin{equation}
\langle O \rangle ={1\over Z} \int {\cal D}\phi {\cal D}\xi \ O(\phi, \xi)\ 
{\rm e}^{-\int d^2{\bf x} \ {q \kappa \over 2\pi} \left( \partial_i \phi \right)^2 + { \kappa\over 8\pi q} \left( \partial_i \xi\right)^2
- 2z {\rm cos} \xi} \ ,
\label{integral}
\end{equation}
where we have renormalized the angle $\xi$ to lie in the interval $[0,2\pi]$ for comparison with 
Gauge field observables in the entangled mixed state are thus determined by the classical partition function of the 2D sine-Gordon model, or equivalently the 2D XY model\,\cite{itzykson}. While the original $\phi$ field plays the role of the spin wave field, the new dynamical sine-Gordon field $\xi$ describes the vortex dynamics. We can now use the classical results on the 2D sine-Gordon (or XY) renormalization group flow\,\cite{itzykson} also here.  

The 2D XY model undergoes the famed Berezinskii-Kosterlitz-Thouless (BKT)\,\cite{bkt} phase transition. The value $\kappa =q/2$ separates a weak-coupling phase for $\kappa < q/2$ from a strong coupling phase for
$\kappa >q/2$.  For the case $q=2$ of Cooper pairs this value corresponds to the self-dual point $\kappa = 1$. 
In both phases the coupling constant $\kappa $ flows to large values in the IR limit. In the weak coupling phase, the fugacity $z \to 0$ in the IR limit while in the strong coupling phase $z\to \infty$ in the IR limit. The renormalization flow depends on a constant $C$, which represents a particular combination of the initial conditions for the flow. In the case $C>0$, of interest here, there is no IR fixed-point in the strong-coupling phase, while the BKT critical point $\kappa_{\rm crit} =q/(2+C) $ is a confining IR fixed point ($z$=0). As in QCD, the coupling constant $\kappa $ flows to small values in the ultraviolet (UV) regime. Contrary to QCD, it is the confining IR fixed point which is perturbative, while the asmyptotically free UV regime is non-perturbative. 

Before proceeding, let us remark that the dual critical point at $\kappa_{\rm crit} = (2+C)/q$ represents the dual superconducting phase, while the intermediate regime $q/(2+C) < \kappa < (2+C)/q$, $z=0$ is the Bose metal phase first predicted in\,\cite{dst} and recently shown to be a bosonic topological insulator\,\cite{bm}. For $C<0$ the BKT renormalization flow implies a direct transition between superinsulator and superconductor. 

Let us now prove that the charge entanglement regime is linearly confining. To this end we first note that it is a condensate phase for vortices. As is evident from (\ref{integral}), near the IR fixed point, the phase $\xi$ of the gauge field has correlations screened on the length $\lambda_{\rm vor}= \sqrt{\kappa/2\pi qz} $, which diverge at the fixed point. These diverging correlations imply diverging fluctuations of the vortex number near the fixed point, the characteristics of a condensate. 
As has been first ponted out in\,\cite{polyakov} this vortex condensate phase in a 2D compact gauge theory is characterized by the presence of instantons. Let us compute thus the effect of instantons on the charge-anticharge potential. To this end we shall consider again the insertion of two external probe charges of different sign by the exponential line integral factor (\ref{ext}), focusing specifically on the effect of the new dynamical field $\xi$. Since this is a phase, at first sight it would seem that this effect reduces simply to two phase factors at the end of the path. Instantons, however, can cause the phase to jump on the path, leading to large effects. 
Following\,\cite{sakita} we write the energy shift caused by single instantons/anti-instantons on the path connecting the two charges as 
\begin{equation}
\Delta E_{\rm inst} = E_{\rm inst}  \int_{\rm path}  d{\bf x}  \left (1 - {\rm Re} \left[ K^{-1} ({\bf x}_1, {\bf x})  K({\bf x}_2, {\bf x}) \right] \right) \ ,
\label{inst1}
\end{equation}
where $E_{\rm inst}$ is the contribution of instantons in absence of external charges and $K({\bf x}_2, {\bf x})$ denotes the exponential of the instanton function valued at ${\bf x}_2$ for an instanton located at some ${\bf x}$ on the path. Since the instanton solution corresponds to a phase jump at ${\bf x}$ we obtain 
\begin{equation}
\Delta E_{\rm inst} = E_{\rm inst}  \int_{\rm path}  d{\bf x}  \left (1 - {\rm cos} \left( Q \Delta \lambda_{\rm inst} ({\bf x}) \right) \right) \ .
\label{inst2}
\end{equation}
In compact U(1) gauge theory, the instantons\,\cite{polyakov} correspond to unit magnetic monopoles in 3D Euclidean space\,\cite{olive}, which implies a magnetic flux $2\pi / q$ on the 3D unit sphere. The corresponding phase jumps at fixed time in Minkowski space-time are thus $\Delta \lambda_{\rm inst} = \pi /q$. This gives the final result
\begin{equation} 
\Delta E_{\rm inst} = E_{\rm inst}  \left (1 - {\rm cos} \left( {\pi Q \over q}  \right) \right) R \ ,
\label{inst3}
\end{equation}
where $R$ is the separation of the charge-anticharge pair. This is Polyakov's classical\,\cite{polyakov} result that instantons in the vortex condensation phase cause linear confinement for charges $Q=q$ satisfying the Dirac quantization condition, while double charges $Q=2q$ are non-confined. 

These results show that 2D QED with compact Chern-Simons dynamical matter is an asymptotically free theory with linear confinement due to the strong entanglement of charge in a vortex condensate. As such it is a single-colour ``toy model" for QCD. The fact that an exact duality mapping to a confining string exists\,\cite{conf} makes it into a complete, exactly solvable model of strong gauge interactions. It is remarkable that this model is explicitly realized in condensed matter as the superinsulating phase of thin superconducting films\,\cite{vinokur, electrostatics}.

\end{document}